\documentclass[a4paper]{article}
\usepackage{amsmath,graphicx}
\usepackage{booktabs}
\usepackage{times}
\usepackage{latexsym}
\usepackage{graphicx}
\usepackage{booktabs}
\usepackage{subfig}
\usepackage[textsize=large]{todonotes}
\usepackage{booktabs}
\usepackage{algorithm}
\usepackage{algpseudocode}
\usepackage[inline]{enumitem}
\usepackage{amsmath,amssymb}
\usepackage[font=small,skip=5pt]{caption}
\usepackage{tabularx}
\usepackage{url}

\newcommand{\param}[1]{\vec{\theta}_{\text{#1}}}

\renewcommand{\vec}[1]{{\boldsymbol{\mathbf{#1}}}}   

\DeclareMathOperator{\R}{{\rm I\!R}}

\usepackage{INTERSPEECH2019}

\title{Pretraining by Backtranslation for End-to-end ASR in Low-Resource Settings}
\name{Matthew Wiesner$^\dagger$, Adithya Renduchintala$^\dagger$, Shinji Watanabe$^\dagger$, \\Chunxi Liu$^\dagger$, Najim Dehak$^\dagger$, Sanjeev Khudanpur$^\dagger$$^\ddagger$ \thanks{This work was supported by DARPA LORELEI Grant N\b{o} HR0011-15-2-0024 and partially carried out during the 2018 Jelinek Memorial Summer Workshop on Speech and Language Technologies, supported by gifts from Microsoft, Amazon, Google, Facebook, and MERL/Mitsubishi Electric.}} 
\address{$^\dagger$Center for Language and Speech Processing, The Johns Hopkins University, USA\\
$^\ddagger$Human Language Technology Center of Excellence, The Johns Hopkins University, USA}
\email{\{wiesner,adi.r,shinjiw,cliu77,ndehak3,khudanpur\}@jhu.edu}

\begin{document}
\maketitle
\begin{abstract}
  We explore training attention-based encoder-decoder ASR in low-resource settings. These models perform poorly when trained on small amounts of transcribed speech, in part because
  they depend on having sufficient target-side text to train the attention and decoder networks. In this paper we address this shortcoming by pretraining our network parameters using only text-based data and transcribed speech from other languages. We analyze the relative contributions of both sources of data. Across 3 test languages, our text-based approach resulted in a 20\% average relative improvement over  a text-based augmentation technique without pretraining. Using transcribed speech from nearby languages gives a further 20-30\% relative reduction in character error rate.
\end{abstract}

\section{Introduction}
\label{sec:introduction}

Attention-based encoder-decoder networks have achieved state-of-the art performance in ASR when trained on over 12k hours of transcribed speech \cite{chiu2018state}, but their performance lags behind conventional systems in more moderate resource conditions and has only just begun to be studied in low-resource conditions \cite{rosenberg2017end,cho2018multilingual}. One way to improve ASR performance without access to more transcribed speech is to leverage linguistic resources from other languages and modalities. Bolstering the decoder with a language model (LM) trained on supplemental text data is one such method that improves end-to-end ASR performance \cite{chorowski2016towards,hori2017advances}; however, more significant gains can be obtained by training on additional synthetically perturbed speech \cite{ragni2014data,ko2015audio,hannun2014deep}, or by \emph{multilingual training}, which augments the training data with transcribed speech from other languages \cite{lin2009study,ghoshal2013multilingual,watanabe2017language,li2017multi,kim2017towards,toshniwal2018end}

Using an LM in decoding is appealing as it only uses text data, but provides only modest improvements in performance. 
And while multilingual training often provides more significant improvements in performance, this approach also requires additional transcribed speech, preferably from similar languages~\cite{watanabe2017language,adams2019massively}.
Our aim is to achieve performance improvements similar to multilingual training, but obtained solely from text data.

As a starting point we  consider multi-modal data augmentation (MMDA): a data augmentation scheme for encoder-decoder based ASR which only requires text data~\cite{renduchintala2018MMDA} (see figure \ref{fig:bothfigures}. (a)). The approach, inspired by ``back-translation'' in neural machine translation (NMT) \cite{Sennrich2016BackTranslation}, involves using an additional \emph{augmenting} encoder (in addition to the traditional acoustic encoder), which accepts a sequence of features derived from text as input and learns to predict the original text. Other work uses a text-to-speech (TTS) system to generate the augmenting features, however, training a reasonable TTS requires more single speaker data than we have available in many low-resource situations \cite{hayashi2018back,mimura2018leveraging}.

We extend MMDA to work in low resource contexts by again borrowing from techniques in NMT. We adapt a technique proposed in \cite{ramachandran2016unsupervised}, which uses an unsupervised language modeling task on both the source and target languages to pretrain the encoder and decoder model parameters respectively in order to reduce over-fitting and improve generalization. This approach uses the insight that in NMT both the encoder and decoder act as language models. In ASR, however, only the decoder clearly exhibits this behavior. Unlike in NMT, however, ASR exhibits monotonic attention, which is relatively easy to initialize. For this reason we instead pretrain the decoder and attention parameters using synthetically ''back-translated`` training examples with MMDA.

To pretrain the encoder we propose a modified architecture that feeds the output of the augmenting encoder to the acoustic encoder. The augmenting data can then be viewed as \emph{pseudo-speech} from some language that we add to our training data. 
We refer to this as pseudo-speech data augmentation (PSDA) as the augmenting encoder is implicitly tasked with learning representations of the augmenting data that resemble the original acoustic features. To study the usefulness the synthetic data in pretraining the encoder, we also compare pretraining the network with PSDA and synthetic inputs to training with actual speech from other languages.

\section{Related Work}
\label{sec:related_work}
The most similar work is \cite{renduchintala2018MMDA}, which uses categorical data in addition to transcribed speech when training the attention and decoder networks in end-to-end ASR. Related work on how to best integrate language models into end-to-end ASR includes deep and shallow fusion \cite{gulcehre2015using}, cold fusion \cite{sriram2017cold}, or transfer fusion \cite{inaguma2018transfer}. While \cite{ramachandran2016unsupervised} proposes pretraining both the encoder and decoder on unpaired monolingual data using an \emph{unsupervised} language modeling objective, we propose using back-translated data as in \cite{renduchintala2018MMDA}, to pretrain using a \emph{supervised} objective.

Work in multilingual (pre)training has the same objective as \cite{ramachandran2016unsupervised} of increasing the generalizability of the encoder. \cite{watanabe2017language} and \cite{toshniwal2018end} have both investigated multilingual training of encoder-decoder architectures. \cite{adams2019massively} showed that pretraining the encoder using data from other, preferably nearby, languages can result in large performance gains in encoder-decoder ASR.

In our work we pretrain both the encoder and decoder, as in  \cite{ramachandran2016unsupervised}. For the decoder we pretrain only on augmenting data using MMDA \cite{renduchintala2018MMDA} instead of a language modeling objective. For the encoder we explore training using speech in other languages as well as a novel architecture, PSDA, to enable joint pretraining of the encoder and decoder on back-translated text.


\section{Data Augmenting Architecture}\label{sec:architecture}

\begin{figure}%
    \centering
    \subfloat[MMDA Architecture]{
\tikzset{every picture/.style={line width=0.75pt}} 

\begin{tikzpicture}[x=0.75pt,y=0.75pt,yscale=-1,xscale=0.75]

\draw    (120,190) -- (80,170) ;
\draw [shift={(80,170)}, rotate = 386.57] [color={rgb, 255:red, 0; green, 0; blue, 0 }  ]   (0,0) .. controls (3.31,-0.3) and (6.95,-1.4) .. (10.93,-3.29)(0,0) .. controls (3.31,0.3) and (6.95,1.4) .. (10.93,3.29)   ;

\draw    (30,190) -- (70,170) ;
\draw [shift={(70,170)}, rotate = 513.4300000000001] [color={rgb, 255:red, 0; green, 0; blue, 0 }  ]   (0,0) .. controls (3.31,-0.3) and (6.95,-1.4) .. (10.93,-3.29)(0,0) .. controls (3.31,0.3) and (6.95,1.4) .. (10.93,3.29)   ;

\draw    (75,150) -- (75,130) ;
\draw [shift={(75,130)}, rotate = 450] [color={rgb, 255:red, 0; green, 0; blue, 0 }  ]   (0,0) .. controls (3.31,-0.3) and (6.95,-1.4) .. (10.93,-3.29)(0,0) .. controls (3.31,0.3) and (6.95,1.4) .. (10.93,3.29)   ;

\draw    (74,111) -- (74,91) ;
\draw [shift={(74,91)}, rotate = 450] [color={rgb, 255:red, 0; green, 0; blue, 0 }  ]   (0,0) .. controls (3.31,-0.3) and (6.95,-1.4) .. (10.93,-3.29)(0,0) .. controls (3.31,0.3) and (6.95,1.4) .. (10.93,3.29)   ;

\draw    (115,230) -- (115,210) ;
\draw [shift={(115,210)}, rotate = 450] [color={rgb, 255:red, 0; green, 0; blue, 0 }  ]   (0,0) .. controls (3.31,-0.3) and (6.95,-1.4) .. (10.93,-3.29)(0,0) .. controls (3.31,0.3) and (6.95,1.4) .. (10.93,3.29)   ;

\draw    (36,230) -- (36,210) ;
\draw [shift={(36,210)}, rotate = 450] [color={rgb, 255:red, 0; green, 0; blue, 0 }  ]   (0,0) .. controls (3.31,-0.3) and (6.95,-1.4) .. (10.93,-3.29)(0,0) .. controls (3.31,0.3) and (6.95,1.4) .. (10.93,3.29)   ;

\draw  [fill={rgb, 255:red, 184; green, 233; blue, 134 }  ,fill opacity=0.3 ]  (75, 190) rectangle (157, 210)   ;
\draw  [fill={rgb, 255:red, 245; green, 166; blue, 35 }  ,fill opacity=0.3 ]  (0, 190) rectangle (70, 210)   ;
\draw  [fill={rgb, 255:red, 80; green, 227; blue, 194 }  ,fill opacity=0.3 ]  (35, 150) rectangle (116, 170)   ;
\draw  [fill={rgb, 255:red, 74; green, 144; blue, 226 }  ,fill opacity=0.3 ]  (35, 110) rectangle (116, 130)   ;
\draw  [color={rgb, 255:red, 255; green, 255; blue, 255 }  ,draw opacity=1 ]  (43, 267) rectangle (113, 294)   ;

\draw (34,200) node  [align=left] {{\scriptsize Enc ($\param{enc}$)}};
\draw (75,160) node  [align=left] {{\scriptsize Attention ($\param{att}$)}};
\draw (73.5,120) node  [align=left] {{\scriptsize Decoder ($\param{dec}$)}};
\draw (115,200) node  [align=left] {{\scriptsize Aug Enc ($\param{DA}$)}};
\draw (75,80) node  [align=left] {$\vec{y}$};
\draw (115,240) node  [align=left] {$\hat{\vec{x}}$};
\draw (36,240) node  [align=left] {$\vec{X}$};

\end{tikzpicture}

  }%
    \qquad
    \subfloat[PSDA Architecture]{
\tikzset{every picture/.style={line width=0.75pt}} 

\begin{tikzpicture}[x=0.75pt,y=0.75pt,yscale=-1,xscale=0.75]

\draw    (82.5, 172.5) rectangle (152.5, 192.5)   ;

\draw    (76, 133) rectangle (158, 153)   ;

\draw    (119, 218) rectangle (200, 238)   ;

\draw    (76, 93) rectangle (158, 113)   ;

\draw    (164,218) -- (129.6,192.6) ;
\draw [shift={(129.6,192.6)}, rotate = 396.44] [color={rgb, 255:red, 0; green, 0; blue, 0 }  ]   (0,0) .. controls (3.31,-0.3) and (6.95,-1.4) .. (10.93,-3.29)(0,0) .. controls (3.31,0.3) and (6.95,1.4) .. (10.93,3.29)   ;

\draw    (117.5,172.5) -- (118,153) ;
\draw [shift={(118,153)}, rotate = 451.47] [color={rgb, 255:red, 0; green, 0; blue, 0 }  ]   (0,0) .. controls (3.31,-0.3) and (6.95,-1.4) .. (10.93,-3.29)(0,0) .. controls (3.31,0.3) and (6.95,1.4) .. (10.93,3.29)   ;

\draw    (117,133) -- (117,113) ;
\draw [shift={(117,113)}, rotate = 450] [color={rgb, 255:red, 0; green, 0; blue, 0 }  ]   (0,0) .. controls (3.31,-0.3) and (6.95,-1.4) .. (10.93,-3.29)(0,0) .. controls (3.31,0.3) and (6.95,1.4) .. (10.93,3.29)   ;

\draw    (116,94) -- (116,74) ;
\draw [shift={(116,74)}, rotate = 450] [color={rgb, 255:red, 0; green, 0; blue, 0 }  ]   (0,0) .. controls (3.31,-0.3) and (6.95,-1.4) .. (10.93,-3.29)(0,0) .. controls (3.31,0.3) and (6.95,1.4) .. (10.93,3.29)   ;

\draw    (159,258) -- (159,238) ;
\draw [shift={(159,238)}, rotate = 450] [color={rgb, 255:red, 0; green, 0; blue, 0 }  ]   (0,0) .. controls (3.31,-0.3) and (6.95,-1.4) .. (10.93,-3.29)(0,0) .. controls (3.31,0.3) and (6.95,1.4) .. (10.93,3.29)   ;

\draw    (78,213) -- (109.2,192.6) ;
\draw [shift={(109.2,192.6)}, rotate = 506.82] [color={rgb, 255:red, 0; green, 0; blue, 0 }  ]   (0,0) .. controls (3.31,-0.3) and (6.95,-1.4) .. (10.93,-3.29)(0,0) .. controls (3.31,0.3) and (6.95,1.4) .. (10.93,3.29)   ;

\draw  [fill={rgb, 255:red, 184; green, 233; blue, 134 }  ,fill opacity=0.3 ]  (119, 218) rectangle (200, 238)   ;
\draw  [fill={rgb, 255:red, 245; green, 166; blue, 35 }  ,fill opacity=0.3 ]  (82.5, 172.5) rectangle (152.5, 192.5)   ;
\draw  [fill={rgb, 255:red, 80; green, 227; blue, 194 }  ,fill opacity=0.3 ]  (76, 133) rectangle (158, 153)   ;
\draw  [fill={rgb, 255:red, 74; green, 144; blue, 226 }  ,fill opacity=0.3 ]  (76, 93) rectangle (158, 113)   ;

\draw (117,63) node  [align=left] {$\vec{y}$};
\draw (159,268) node  [align=left] {$\hat{\vec{x}}$};
\draw (78,223) node  [align=left] {$\vec{X}$};
\draw (115.5,103) node  [align=left] {{\scriptsize Decoder ($\param{dec}$)}};
\draw (159,228) node  [align=left] {{\scriptsize Aug Enc ($\param{DA}$)}}; 
\draw (117,143) node  [align=left] {{\scriptsize Attention ($\param{att}$)}};
\draw (116.5,182.5) node  [align=left] {{\scriptsize Enc ($\param{enc}$)}}; 

\end{tikzpicture}

  }%
    \caption{MMDA and PSDA. $\vec{X}$ is the original speech, $\vec{\hat{x}}$ is the text-based augmenting input, and $\vec{y}$ is an output character sequence. $\param{enc}$, $\param{att}$, $\param{dec}$, $\param{DA}$, are the parameters corresponding to the encoder, attention, decoder, and data-augmenting encoder respectively}
\label{fig:bothfigures}%
\end{figure}
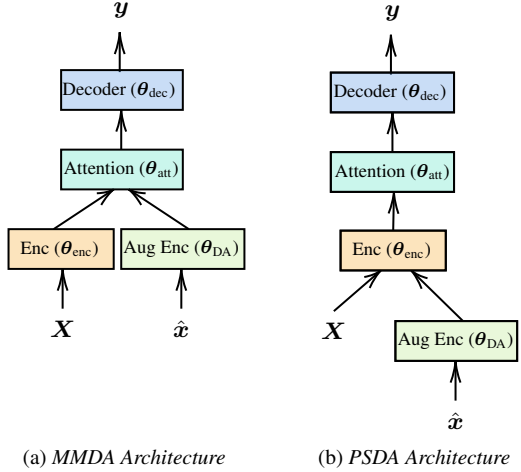

Our architecture follows the encoder-decoder model which maximizes the log-likelihood:
\begin{align}
  \mathcal{L}(\vec{\theta}) & = \log P(\vec{y} \mid \vec{X}\; ; \param{enc}, \param{att}, \param{dec})
\end{align}
$\vec{y}$ denotes the desired output character sequence and $\vec{X} \in \R^{L \times D}$ a tensor of speech frames of length $L$ and feature dimension $D$. We denote the entire set of network parameters by $\vec{\theta}$, which is composed of acoustic encoder parameters $\param{enc}$, attention mechanism parameters $\param{att}$ and decoder parameters $\param{dec}$.
The encoder consists of a projection-biLSTM  with a pyramidal structure for the acoustic encoder~\cite{chan2016listen}, the decoder is a single-layer LSTM and location-aware attention completes the entire end-to-end network.
\subsection{Multi-Modal Data Augmentation}
The MMDA technique (fig. \ref{fig:bothfigures}a) transforms the objective into a multi-task objective:
\begin{align}
\small
  \mathcal{L}(\vec{\theta}) = \begin{cases}
    \log P(\vec{y} \mid \vec{X}\; ; \param{enc}, \param{att}, \param{dec})\;\text{speech} \\
    \log P(\vec{y} \mid \hat{\vec{x}}\; ; \param{DA}, \param{att}, \param{dec})\;\text{text-based}
  \end{cases} \label{eq:mmda_obj}
\end{align}
When the inputs are acoustic features, $\vec{X}$, MMDA uses the standard encoder-decoder network to maximize the original ASR objective and the primary task.
When the inputs are text-based features, $\hat{\vec{x}}$, MMDA uses a \emph{data-augmenting encoder} instead of the acoustic encoder and maximizes the probability of the output sequence paired with the text-based representation (secondary task).
In Eq~\ref{eq:mmda_obj}, $\param{DA}$ denotes the parameters of the data-augmenting encoder which is composed of an embeddings layer and a single-layer projection-biLSTM.
\subsection{Pseudo-Speech Data Augmentation}
We propose a variation of MMDA which changes the architecture during the secondary task (fig. \ref{fig:bothfigures}b).
In this setup, we cascade the data-augmenting and acoustic encoders and force the data-augmenting encoder's output to match the dimensionality of acoustic frames (which are the input in the primary task).
Thus, in PSDA the entire encoder-decoder network is part of the computation graph in both tasks.
\begin{align}
\small
  \mathcal{L}(\vec{\theta}) = \begin{cases}
    \log P(\vec{y} \mid \vec{X}\; ; \param{enc}, \param{att}, \param{dec})\;\text{speech} \\
    \log P(\vec{y} \mid \hat{\vec{x}}\; ; \param{DA}, \param{enc}, \param{att}, \param{dec})\;\text{text-based}
  \end{cases} \label{eq:psda_obj}
\end{align}
PSDA can be viewed as a proxy multilingual training method, where the pseudo-speech generated by the data-augmenting encoder (which is fed into the acoustic encoder) is a cheap approximation of real acoustic features of some new, but related language.
We use the same structure for the data-augmenting encoder as in the MMDA case.

\subsection{Multi-task Training \& Pretraining}
We pretrain our encoder and decoder with augmenting data using both MMDA and PSDA architectures and show that it significantly improves ASR performance. It is important to occasionally update the network using augmenting data after pretraining in order to prevent catastrophic forgetting \cite{goodfellow2013empirical}. 

In \cite{renduchintala2018MMDA} we proposed training the MMDA network by alternating between audio-data and augmenting-data minibatches. We now allow for more flexibility by using a hyper-parameter $\rho \in (0,1)$ that decides if the model should be trained on speech data or text-based data. In this way we can tune the number of augmenting updates needed to prevent catastrophic forgetting.

\section{Shallow Fusion}
We compare MMDA, PSDA, and our pretrained variants to a shallow fusion baseline. Shallow fusion \cite{gulcehre2015using}, is a simple, effective and commonly used technique for external language model integration in sequence to sequence learning for ASR \cite{toshniwal2018comparison}. In shallow fusion, a list of partial hypothesis and corresponding scores is produced by the ASR decoder. Each partial hypothesis is then also scored by an external language model. A composite score for the partial hypothesis is given by
\begin{equation}
    \mbox{score}(\vec{y}) = \log{P_{ASR}(\vec{y} | \vec{x})} + \lambda \log{P_{LM}(\vec{y})}.
\end{equation} $\log{P_{ASR}(\vec{y} | \vec{x})}$ is the ASR score for a hypothesis sequence $\vec{y}$ given an input utterance $\vec{x}$, $\log{P_{LM}(\vec{y})}$ is the corresponding language model score, and $\lambda$ is a tunable parameter. The list of hypotheses is reordered prior to prediction of the subsequent output and only the top scoring hypotheses are retained. \cite{renduchintala2018MMDA} showed that shallow fusion and MMDA result in similar, but complimentary performance gains. The advantage of MMDA is a simplified decoding strategy.

\section{Experiments}
We conducted experiments on 4 languages from the Voxforge corpus: Catalan, Portuguese, Italian, and French. We chose these data sets because they have small amounts of relatively clean training data (0.5-30h) and are closely related to Spanish which we used in multilingual training (see \ref{sec:multiling}). This allows us to study the effect of small training data on end-to-end ASR in isolation, without worrying about confounding factors such as language relatedness or the noisiness of the training data.

For Catalan, Portuguese, and Italian,  we created 5 baseline systems:
\begin{enumerate*}
    \item A baseline monolingual model (Monolingual) 
    \item A monolingual model with decoded using shallow fusion (LM)
    \item The same baseline model trained as in \cite{renduchintala2018MMDA} using an augmenting encoder and augmenting data scraped from the web (MMDA).
    \item A model that was trained on transcribed speech from other languages in addition to the monolingual data (ML).
    \item The multilingual model decoded with shallow fusion (ML+LM).
\end{enumerate*}
All of the augmenting data was used to train the RNNLM for each language to enable a fair comparison between shallow fusion and MMDA.

\subsection{Monolingual Systems}
\label{sec:monoling}
We trained monolingual systems for Catalan, Portuguese, and Italian. The training, development and evaluation sets are constructed by randomly sampling 80\%, 10\%, and 10\% of the data for each set respectively, ensuring that no prompt in the development or test sets is duplicated in the training set. The Catalan and Portuguese systems were trained on the entire 30 min and 3 hour extracted training sets respectively. For Italian we trained only on a 4 hour subset of the full 16 hour training set in order to more closely mimic the training conditions of the two other languages. All systems were trained using ESPnet \cite{watanabe2018espnet}. We trained encoder-decoder networks as described in \ref{sec:architecture}, but without the augmenting encoder. We used the same configurations as in \cite{watanabe2017language}, except for we  used 4 encoder layers for all experiments.

\subsection{Multilingual Systems}
\label{sec:multiling}
For Catalan and Portuguese we augmented the training data with all 30h of the Hub-4 Spanish Broadcast news corpus training set and all 16h of the Italian Voxforge training set. For Italian we only added the Hub-4 Spanish to training. All systems were trained using the same network configurations and training parameters as the monolingual systems. We followed \cite{toshniwal2018end,watanabe2017language} and use as output symbols the union of all graphemes seen in training such that the network was capable of outputting any of the languages seen in training.

\subsection {MMDA \& PSDA}
\label{sec:MMDA+PSDA}
We trained monolingual MMDA and PSDA systems as well as systems with pretraining (MMDA+P, PSDA+P) as described in section \ref{sec:architecture} using the same data splits as described in section \ref{sec:monoling}. 
We also trained multilingual (ML) MMDA and PSDA systems which we compared to a ML baseline with RNNLM shallow fusion (ML+LM).
\begin{table}[ht!]
\small
\centering
\caption{Summary of \textbf{monolingual} experiments. We see that our proposed pretraining (indicated with {\bf +P}) improves performance dramatically. Both MMDA+P and PSDA+P show strong and consistent improvement over Monolingual, LM and MMDA baselines, reducing CER by 20\% to 26\%.}
\begin{tabular}{@{}lccc@{}}
\toprule
\textbf{Task} & \begin{tabular}[c]{@{}c@{}}CA (0.5h) \\dev, eval\end{tabular} & \begin{tabular}[c]{@{}c@{}}PT (3h) \\dev, eval\end{tabular}  & \begin{tabular}[c]{@{}c@{}}IT (4h) \\dev, eval\end{tabular} \\ \midrule
\textbf{Monolingual} & 85.2, 82.3 & 76.9, 80.1 & 31.2, 31.4 \\
\textbf{LM} & 79.7, 76.9 & 77.6, 79.9 & 32.1, 32.1 \\
\textbf{MMDA} & 79.1, 76.5 & 73.7, 72.3 & 27.9, 28.2 \\
\midrule
\textbf{PSDA} & 86.3, 81.4 & 80.0, 76.9 & 29.2, 29.4 \\
\textbf{MMDA + P} & 73.8, 75.3 & 55.4, 56.1 & \textbf{23.9, 24.1} \\
\textbf{PSDA + P} & \textbf{71.2, 72.2} & \textbf{47.4, 50.2} & 25.0, 26.0 \\
\bottomrule
\end{tabular}
\label{tab:lowrec}
\vspace{6mm}
\caption{Summary of \textbf{multilingual} experiments (indicated with ML). MMDA+P and PSDA+P yield performance gains beyond multilingual training and RNNLM fusion for both PT and IT.}
\small
\centering
\resizebox{\linewidth}{!}{
\begin{tabular}{@{}lccc@{}}
\toprule
\textbf{Task} & \begin{tabular}[c]{@{}c@{}}CA (0.5h)\\ dev, eval\end{tabular} & \begin{tabular}[c]{@{}c@{}}PT (3h) \\ dev, eval\end{tabular}  & \begin{tabular}[c]{@{}c@{}}IT (4h)\\ dev, eval\end{tabular} \\ \midrule
\textbf{ML} & 33.1, 37.2 & 34.5, 38.4 & 20.1, 21.0 \\
\textbf{ML+LM} & \textbf{31.1}, 36.4& 33.3, 37.7 & 18.7, 19.6 \\
\midrule
\textbf{MMDA+P+ML+LM} & 34.2, \textbf{36.2} & \textbf{32.4}, 35.9 & 17.2, 17.8 \\
\textbf{PSDA+P+ML+LM} & 34.9, 38.7 & 33.8, \textbf{35.3} & \textbf{17.1, 17.6} \\
\bottomrule
\end{tabular}}
\label{tab:MLTable}
\vspace{3mm}
  \centering
  \caption{Example VoxForge Italian sentence (criptogenetico-criptogenetico-20081224-jmd-it-0801) decoded using 4 ASR systems trained on 4h of speech. LM is the baseline system decoding using langauge model shallow fusion. PSDA+P refers to PSDA with pretraining. ML+LM is multilingual training and language model shallow fusion. COMB uses all techniques above combined. The development set word-error-rate (WER) of each model is shown. Results on the corresponding evaluation sets are always 2-3\% worse. Word errors are \textbf{bold}.}
  \resizebox{\linewidth}{!}{
  \begin{tabular}{lll}
    \hline
    \textbf{System} & WER & \textbf{Sentence} \\ \hline
   LM & 74.9
   & \texttt{\uppercase{queste \textbf{SEI VERSO NON} nostre}}
   \\ & &
   \texttt{\uppercase{\textbf{TORNATO} I quindi \textbf{NOSTRI PER SI}}}
   \\ & & \texttt{\uppercase{\textbf{ERANO DI BANDERE A} tutti}} \\\midrule[0.05pt]
    PSDA+P & 70.1
    & \texttt{\uppercase{queste \textbf{SE IL VESO  NON  MOSTRE}}}
    \\ & &
    \texttt{\uppercase{tornate quindi \textbf{E}  vostri \textbf{PENSI}}}
    \\ & & \texttt{\uppercase{\textbf{E} noi \textbf{DI} mangeremo \textbf{A} tutti}} \\ \midrule[0.05pt]
    ML+LM & 60.7
    & \texttt{\uppercase{queste selve \textbf{SON A} nostre}}
    \\ & &
    \texttt{\uppercase{tornate \textbf{QUINDIE NOSTRI} paesi}}
    \\ & & \texttt{\uppercase{\textbf{E} noi \textbf{DI} mangeremo tutti}}
   \\ \midrule[0.05pt]
    COMB & 56.2
    & \texttt{\uppercase{queste selve sono nostre}}
    \\ & &
    \texttt{\uppercase{tornate quindi \textbf{E}  vostri \textbf{PESI}}}
    \\ & &
    \texttt{\uppercase{\textbf{E} noi \textbf{DI} mangeremo tutti}} \\ \hline
    Reference & & \texttt{\uppercase{queste selve sono nostre}}
    \\ & &
    \texttt{\uppercase{tornate quindi AI vostri PAESI}}
    \\ & &
    \texttt{\uppercase{O noi VI mangeremo tutti}} \\ \hline
  \end{tabular}
  }
  \label{tab:example_outputs}
\end{table}

\begin{figure*}[th!]
    \includegraphics[width=\linewidth]{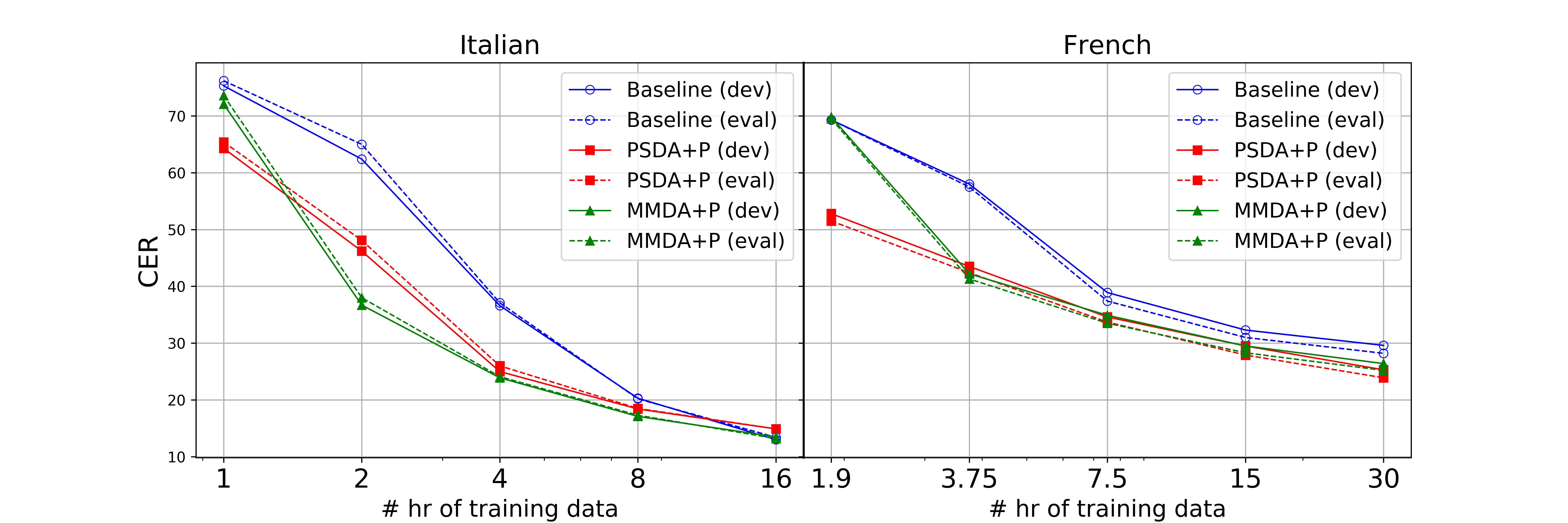}
  \caption{CER of the baseline system, MMDA+P, and PSDA+P on the Voxforge Italian and French Corpora across varying training set sizes}
  \label{fig:trainamounts}
\end{figure*}

\noindent {\bf Augmenting Data:} 
The augmenting data were generated by first scraping Wikipedia for text in the language of interest. We then filtered out tokens with characters that did not appear in the audio training data as well as long and short sentences, resulting in 2.2, 3.8, 3.2, and 4.2 million training examples for Catalan, Portuguese, Italian, and French respectively. As in \cite{renduchintala2018MMDA} we converted this text into sequences of phonemes which was shown to give better performance than simply using only the graphemes. We created pronunciation lexicons for each language by scraping Wiktionary for pronunciations of all words seen in the augmenting text data. For each language we then trained a grapheme-to-phoneme transducer using \texttt{Phonetisaurus} \cite{novak2012wfst} on the corresponding scraped lexicons, which we then used to recover pronunciations for all words in the augmenting data absent from the lexicon.

\cite{renduchintala2018MMDA} also found that modeling phoneme duration was important, and use the frame level phoneme alignments from TIMIT to learn this model. Transferring the duration model to a new language required manually mapping the new phoneme inventory to TIMIT phonemes. We instead model phoneme duration with a shared Gaussian distribution across all phonemes, whose mean is the average ratio of input frames to output symbols in the audio training data. We then repeated each phoneme by a duration sampled from this distribution. The variance ensured that any duplicate or similar sentences resulted in unique training pairs. This technique eliminated the need for frame-level phoneme alignments. We suspect that using a more sophisticated duration model would result in better performance, however, this method was simple, effective, and broadly applicable to any language.

\noindent {\bf Hyper-parameter Optimization:} We randomly sampled hyper-parameters from the possible configurations of \# pretraining batches and augmenting ratio: $\{2000, 5000, 8000\} \times \{0.1, 0.2, 0.5\}$. We selected the parameters that performed best on the development set for each experiment. For the monolingual French and Italian experiments, however, we simply used 2000 pretraining batches and 0.1 and 0.5 augmenting ratios for PSDA and MMDA respectively as we found these values worked well for Portuguese and Catalan.

\section{Results}
Tables \ref{tab:lowrec}, \ref{tab:MLTable} show the performance (CER) of data augmentation across different languages with similar data sizes. We note that vanilla MMDA outperformed shallow fusion (LM) in all 3 languages. The pretrained variants resulted in a further 20\% relative improvement. PSDA+P outperformed pretrained MMDA+P for both Catalan and Portuguese, which have extremely limited training sets, but MMDA+P was the best system on Italian. This corroborated our intuition that PSDA should help more when fewer data are available, as it allows for encoder pretraining, though its utility may only be in extremely data constrained situations. 

We also studied data augmentation on a single language across various amounts of training data. To this end we created 4 smaller Italian and French training sets by successively randomly removing half of the training examples from the original 16 and 30 hour training sets respectively. We then trained the baseline monolingual, MMDA+P, and PSDA+P systems on each resulting dataset using the same network and training parameters as before. We used the same hyperparameters as in the monolingual 4h Italian experiments. Both MMDA+P and PSDA+P performed similarly to each other across all training data sizes, except when training on just a few hours of speech (see fig.\ref{fig:trainamounts}). They both outperformed the baseline by a wide margin, with greater improvements when data were more scarce.

Finally, comparing the use of pseudo-speech features (PSDA+P)  to multilingual training we see that PSDA+P gives about 50\% of the improvement of multilingual training on extremely close languages. Furthermore, the \{MMDA,PSDA\}+P+ML+LM systems were our best performing on the evaluation set in every language tested. Using these techniques together on only 1/4 of the full Italian training data gives performance similar to the baseline model trained on the full data set.

Since pretrained PSDA (PSDA+P) did not outperform pretrained MMDA (MMDA+P) we conclude that most of the gain likely comes from pretraining the decoder and attention parameters. However, since training on other languages seems to help the encoder, we conclude that it is likely the synthetic data itself, and not necessarily PSDA, which is of limited use for pretraining the encoder.

Finally table \ref{tab:example_outputs}. shows the WER of Italian ASR systems and a sample decoded sentence. Appropriate pretraining of the encoder and decoder reduced the WER by 20\% absolute in the 4h Italian set, to 56.2\%. This performance has been shown to still be usable for some downstream tasks such as topic identification in low-resource settings \cite{wiesner2018automatic}.

\section{Conclusion \& Future Work}
We have presented a new data augmentation scheme, PSDA, and demonstrated that pretraining on augmenting data for both MMDA and PSDA outperforms the monolingual, vanilla MMDA and RNNLM shallow fusion baselines. We have shown that without using any additional transcribed speech in any language we can achieve performance improvements approaching those of multilingual training on related languages. Furthermore, our MMDA and PSDA variants improve upon multilingual systems for Portuguese and Italian. Future work should expand upon PSDA by attempting to more explicitly generate speech like features from text, possibly using a generative adversarial network to encourage the augmenting encoder in PSDA to act as a light-weight TTS engine.

\bibliographystyle{IEEEtran}

\bibliography{mybib,library}


\end{document}